ARTSCIENCE (AS) IN CLASSICAL AND QUANTUM PHYSICS

S. L. Weinberg[1]

Department of Physics, Academy of Artscience, Box 1, 18 Langslow Street, Rochester NY 14620-2928

ABSTRACT

A general methodology and specific formalism are used to restrict the Copenhagen interpretation of quantum mechanics. A natural psi-collapse to reality is developed in an equation with terms independent of the measuring equipment.

The theory requires one of three experiments for the ensemble average of position, momentum, or energy, or a probability-experiment.

Both classically and quantum mechanically, we define Artscience (AS) as Logic-Epistemology in Physics, where epistemology is treated as phenomenology. AS (logic-phenomenology) is thus shown to be related to Physics (formalism-data) .

[1] docweinberg@cal.berkeley.edu

1. INTRODUCTION. Artscience (hereinafter AS) - what is it? The answer to this question will be given in this paper from a physical (and scientific) viewpoint. AS is "Logic-Epistemology in Physics", in general holding classically and quantum mechanically (both are treated with examples). In practice, the field of Philosophy also contains metaphysics as what would be a third main branch, but AS does not; AS is explicitly just Logic-Epistemology (logic-phenomenology).

In Sect. 2 we define AS. The discussion of analogies between Physics, the Scientific Method, and AS follows and is summarized in a Table. AS is incomplete, and this is handled. Sect. 3 briefly deals with AS and classical physics.

In Sect. 4, we continue the discussion of AS, treating the Copenhagen interpretation of quantum mechanics and attendant metaphysics. We attempt the derivation of a truly non-ontological (AS) interpretation of quantum mechanics, as a *theory of measurement* containing a *theoretically physical* psi-collapse equation. This physics comes somewhat far into the paper, *i.e.*, in Sect. 4, following necessary preliminary material. The technical and interested reader is directed to read Sect. 4 first and then to read the entire paper subsequent to this for the valuable perspective of a viable non-ontological methodology. Sect. 5 gives the Conclusions.

2. DEFINITIONS AND DISCUSSION. We now narrow-in on what is meant by AS by displaying Definitions 1 - 3 in the Table. We have already emphasized that Def. 3 is the one we propose in this paper.

| 1. PHYSICS | THEORY<br>  a) Postulates<br>  b) Derivation<br>  c) Predictions | EXPERIMENT<br>  a) Empirical state-of-the-art<br>  b) Measurement |
|---|---|---|
| 2. SCIENTIFIC METHOD | HYPOTHESES<br>  a) Mathematical theory<br>  b) Predictions | TEST<br>  a) Experiment<br>  b) Testing hypotheses |
| 3. ARTSCIENCE (AS) | LOGIC<br>  a) Scientific propositions<br>  b) Inference<br>  c) Deductions | EPISTEMOLOGY<br><br>  Physical phenomenology |

Table. Definitions 1, 2, and 3 are compared. The Table is read left-to-right (rows) and up-and-down (columns).

The Table is discussed in the remainder of this Section. Under the main heading in each box in the Table comes a description which are self-explanatory. The breakdown of a Definition in the Table is shown in Columns 2 and 3: Row 1 is the definition (breakdown) of Physics, *etc*. Respective boxes are basically interrelated as we shall see.

As pertains to Def. 1, Max Planck has said that "Physics at its highest level is art." Hereinafter, we take the opposite approach (science vs. art), letting AS be in part different from physics and seeing with what we can identify it. The prefix "Art-" in "Artscience" means then that AS is incomplete or as such "less technical" than physics (physical science). We demand consistently, however, that AS remains technical as far as it holds (technical in the sense of physics), and therefore it is identified as technical-philosophy in physics. (Logic-Epistemology). Aside from Planck's comment and dubbing with the "state-of-the-art", art is obviously less technical and technological than science - so is AS, as we shall see when we study its inherent incompleteness (insufficiency).

Rows 1 and 2 of the Table are standard, complete and filled-in already. We have however a place to put AS in the Table. Note that it is in Column 1, replacing the unwieldy phrases "technical-philosophy in physics" and "Logic-Epistemology in Physics" with AS. Def. 3 is perfectly appropriate and reasonable. It is substantial, accurate, consistently justified, and not merely convenient. We make the consistent identification of AS with Logic-Epistemology throughout this paper.

As for Column 2, scientific propositions are compared directly with hypotheses, and these relate analogously to postulates. Inference (deductions), mathematical theory (predictions), and derivation (predictions) are also interrelated. Column 3 likewise follows naturally (note that any controls are included in a Test).

We are focusing on Def. 3 as the definition of AS. Def. 3 for AS treats physics nominally as logic and phenomenology, the latter being based on data from experimentation (see below; it is understood that not all experiments are state-of-the-art).

Now we have the first of two essential points to make: *Logic is less technical* than formalism in theory in that any derivation in theoretical physics typically relies upon one or more branches of mathematics *other than logic* (it is incomplete, as mentioned above). Logic comprises the hypothetical propositions, the reasoning between equations, the equations themselves (derived from the propositions in a formal sense), and the inferred final result: using mathematics.

As for the third Column of Def. 3, the idea that epistemology is phenomenological is standard ('things themselves'), *q.v.*, the philosophies of Husserl, Heidegger and others (see Sect. 3 for clarification; *cf.* also Mach's phenomenalism, and Bertrand Russell). In physics, we take phenomenology to be an integral part of *experiment*, specifically data-taking, the analysis of data, and error analysis (*cf*. Ne'eman 1998).

Our second essential point may now be made: *Phenomenology is less technical* than experiment because it starts with taking data and is not involved in, say, the design, building or operation of an experiment (again, the incompleteness mentioned above; this comes directly from the definition of "phenomenology"). An example of epistemology in physics is the Copenhagen interpretation (Omnes 1994, p. 85) [see Sect. 4 for problems with this].

The rows of the Table are completely reasonable (necessary) while the columns depict the close relationship of the entries. Since Column 2 entries are all related and similarly for Column 3, Column 1 entries are thus clearly interrelated. In each row, the Column 2 entry is posited, and Column 3 is used in verifying it experimentally or not (exact science). In the terminology of Omnes (1994, p. 529), epistemological theory of knowledge is the correlation of a representation of reality (phenomenology) and logic-mathematics. Then knowledge (AS, a physical law) obtains as the result of their correspondence. We just write this, in our terminology, as Logic-Epistemology (Epistemology in the Table is actually "phenomenology" plus this comparison with logic). The preliminaries for AS have now been completed in this Section.

It should be noted that in this paper we concentrate on the predictive Scientific Method, although not to the exclusion of postdiction (transcendental phenomenology). In practice, in some cases, we may deal with postdictive *models*, or experimental discoveries which have not been predicted, or observations: This is understood.

3. THE ARTSCIENCES IN CLASSICAL PHYSICS. In Sect. 2, we developed AS as Logic-Epistemology (in Physics). Let us now pose the question "What are the artsciences?" [trivially, we are free to use a lower case convention, *e.g.*, for the plural (artsciences), or a specific artscience (logic or epistemology), *etc*.]. By recalling Planck's comment, if taken literally, AS would not merely be technical-philosophy in physics (i.e., logic-phenomenology), but AS would be basic physical science. Instead, as before, we make the obvious and new case that the artsciences are precisely *logic* and *epistemology* in physics (physical science). This is a natural identification and appropriate terminology.

Let us now look at metaphysics in some detail, such as it is philosophy and *not* an artscience in classical physics.

Firstly, we cite a concrete classical example. The reader may note that cosmology has traditionally been metaphysical, but in modern physics it is treated as a mature physical science: This is progress in the history of physics. Anyway, one example of 'persistent' metaphysics cited up to the present is the teleological strong anthropic principle, which we reject. Delsemme (1998) characterizes this, in short, by saying that the universe is explained "in terms of a final purpose: the emergence of humankind." He continues that this principle does not "allow scientific arguments to fit in with metaphysical considerations." We do not subscribe to this blatant myth (which falls outside of Logic-Epistemology).

Secondly, we give another example, still in classical physics, namely Shimony's (1978, p. 12) treatment of Kant's "non-ontological philosophy". This is transcendental philosophy central to which is that we have "no knowledge of things in themselves, but only knowledge of phenomena". Comparing this to phenomenological artscience in Sect. 2, it is seen to be Omnes' "representation of reality", so AS is consistent with non-ontological philosophy (*vs*. metaphysics).

These two examples support our position for AS in the classical case. Thirdly, from Mach's positivist-antimetaphysical philosophy of science to Kant to AS, metaphysics is undesirable and expendable as it is inconsistent and incompatible with science. We leave the classical case here; our following treatment of quantum mechanics, where the problem is, will be much more in depth.

4. AS IN QUANTUM PHYSICS. As mentioned above, AS plays a general role in both classical and quantum physics. We turn our attention now to quantum mechanics in this Section. With our reasoning above intact, we ask what a fully non-ontological AS interpretation of quantum mechanics would say differently than the Copenhagen interpretation about the foundations of quantum mechanics? It turns out that it is equally important to ask 'what it does *not* say (*viz.*, metaphysics)'. We can find ontology mentioned in the literature explicitly in connection with the Copenhagen interpretation, but we consider this non-technical-*philosophy* in physics and, again, *not* AS (Shimony 1978, pp.11-12).

In the literature, there are criticisms (minority view) of the Copenhagen or standard interpretation, but we will not enter into the vagaries (or spectacular successes) of quantum mechanics in this paper beyond treating them somewhat simply, notably the complex $\psi$-function collapse or reduction and the "measurement problem" in the foundations of quantum mechanics. AS is not claimed to be a complete replacement for the Copenhagen interpretation, obviously because of its phenomenological incompleteness (Sect. 2), and yet it is intriguing to see how far we can take this concept with a theory of measurement.

The ideas of Bohr, Heisenberg, and Born say that there is no reality until a measurement is made. This is somewhat analogous to AS/phenomenology as we have derived it, from a totally new perspective, by denying that metaphysics is an artscience and identifying AS with Logic-Epistemology. *Phenomenology* and the Copenhagen interpretation seem roughly concordant and consistent, which is totally unintentional, hence Omnes' phraseology "Copenhagen epistemology". Neither AS/phenomenology nor the Copenhagen interpretation treat reality pre-measurement. We reiterate this point, that AS/phenomenology (a representation of reality) and the Copenhagen interpretation share in common the gross feature of not treating reality before measurement (*cf.* logic-mathematics below); *q.v.*, the Copenhagen interpretation's "contextual reality" or "observation-created reality" (Hobson 1999).

We take 'until a measurement is made' to mean at the end of psi-collapse (or interference). In our terms, measurement begins theoretically when a quantum system is disturbed by some apparatus (the detector) during measurement, *i.e.*, at the point where the respective Hilbert spaces are coupled as

$$\mathcal{H} \equiv H_{system} \otimes H_{apparatus} \quad . \qquad \text{Eq. (1)}$$

Starting with Eq. (1), we attempt to produce a theoretical example of a non-ontological AS interpretation consistent with experiment. Phenomenology alone, starting only with the data-taking stage of experimentation, is not going to completely replace the Copenhagen epistemology. They play different roles in quantum physics, but neither treat what happens before experimental measurement except in theory. AS/logic has however something to add to this.

For the moment we revert to the uncoupled case. When extended, the 'Copenhagen interpretation' *mixes in* theory with what is supposed to be a "contextual" Copenhagen epistemology ("experimental context"). This is done by introducing probability, *viz.*, the wave mechanical probability amplitude $\psi \equiv \psi(x_i, t)$ and the probability density $\mathcal{P} \equiv |\psi|^2$. It claims the status of a quantum system is one of possibility, indeterminate (probabilistic) before it is disturbed as a result of a "contextual" measurement. Instead, the present interpretation is *theoretically physical* (non-ontological) and so are all the equations in it.

Continuing with our example of an ideally non-ontological AS interpretation, we consider a quantum system for which $\psi \in H_{system}$, theoretically, the related absolute value squared being given by the basic step

$$\mathscr{P} = \psi^* \psi \quad . \qquad \text{Eq. (2)}$$

We take $\psi$ to be normalized. The magnitude squared is real and has *physical* dimensions of length$^{-3}$, to wit, the dimensionless probability

$$\Pi(t) \equiv {}_V\!\!\int \mathscr{P}(x_i, t) \, d^3x$$
$$= {}_V\!\!\int |\psi|^2 \, d^3x \quad , \qquad \text{Eq. (3)}$$

which is the probability that a particle is in a volume V at time t (as a function of time; *cf.* wave-particle duality).

$\psi$ is theoretically physical and describes the probabilistic possibilities, as the Copenhagen interpretation implies. It has physical units (see Eq. (3)). $\psi$ is also, naively, a function of a real spacetime coordinate argument solving a partial differential equation, such as the time-dependent Schroedinger equation, for example, with spacetime partial derivatives. All this argues for theoretically physical formalism (not ontology), applicable predictably prior to 'Copenhagen observation'.

Integrating Eq. (3) over all space, we find the system exists with a probability of unity ( $\Pi = 1$; certainty). Rather than saying objective or physical reality exists pre-measurement in quantum mechanics, we shall say that a (degree of) certainty exists in the theory of measurement normalization. This is not the EPR reality criterion (Einstein *et al.* 1935) in another guise, because the system *is* disturbed by measurement.

The complex conjugate plays its usual role as integration using the probability amplitude gives the real expectation value of a dynamical variable (observable), with Hermitian operator $\mathcal{O}$, namely

$$<V> = \int_{\text{all space}} \psi* \mathcal{O} \psi \, d^3x \quad , \qquad \text{Eq.(4)}$$

integrating over all space. N.B.: In brackets <V>, V is a variable and not volume. Eq. (4) models statistical measurement in theory - as a *collapse equation*. $\mathcal{O}\psi$ could be viewed heuristically (at first glance) as something like a collapsed wavefunction, although it is not one (for one thing it has the wrong units). Evidently, this is an effective operator-state-collapse, where operating with $\mathcal{O}$ on $\psi$ and multiplying by $\psi*$ yields psi-collapse (statistically). Below we will incorporate the effects of interference into the collapse. From psi-collapse occurring in Eq. (4), we have obtained the prediction (model) of <V>; Eq. (4) is of course well-known. It is the collapse interpretation that is new. The traditional term "collapse" may be misleading, particularly if $\mathcal{O}\psi$ just gives back $\psi$ times an eigenvalue: Nevertheless we use it, substituting formalism for ontology; we will call $\psi* \mathcal{O} \psi$ the collapse (see Eq. (3) as the collapse probability). Incidentally, probabilities and expectation values obtain in *quantum logic*, which we do not treat here.

State function $\psi$ specifies the state, *i.e.*, it tells us all that we can know about the system. Phenomenology picks up an experiment at the data-taking stage and so says nothing about psi-collapse or "contextual reality". Quantum theory is where we must look for a solution (an understanding of quantum mechanics), as a theory of measurement subject and open to experimental verification, and not as a Copenhagen epistemology.

Theoretically, we have partially solved the problem of how a probability amplitude predicts statistical measurements in reality through Eqs. (3) and (4). An AS interpretation without ontology is useful even if not complete. We may extend our rule of AS to the logic ('no metaphysics', or in this case no ontology) and rely on theory and experiment to give the coupled physics (ff.). We would expect, analogous to Eqs. (2) and (3), to encounter a coupled term

$$P \equiv |\psi + \phi|^2$$
$$= |\psi|^2 + \phi^* \psi + \psi^* \phi + |\phi|^2 \quad , \qquad \text{Eq. (5)}$$

where $\phi \equiv \phi(x_i, t) \in H_{apparatus}$ (the detector) interferes with $\psi \rightarrow \psi + \phi$ predominately in a volume

$$\Delta x \, \Delta y \, \Delta z \quad , \qquad \text{Eq. (6)}$$

having chosen spatial coordinates. In averaging statistically to get these uncertainties, the superposed total wavefunction $\psi + \phi$ is used (first we normalize $\psi$ and $\phi$ and then $\psi + \phi$, this being understood). We note that even uncertainties are real and have physical dimensions, as above, *i.e.*, they are physical quantities (as is the quantum of action).

Eq. (4) in our theory of measurement is simple and basic. A full (and coupled) treatment has to look at an interference-collapse equation integrand like

$$O \equiv (\psi^* + \phi^*) \, \mathcal{O} \, (\psi + \phi)$$
$$= \psi^* \mathcal{O} \psi + \phi^* \mathcal{O} \psi + \psi^* \mathcal{O} \phi + \phi^* \mathcal{O} \phi \qquad \text{Eq. (7)}$$

(analogous to Eq. (4)), where collapse is attributable to $\mathcal{O}$, or to interference cross-terms, or evidently both, *etc.* The factor $(\psi^* + \phi^*) \, \mathcal{O} \, \phi$ in Eq. (7) for O is arguably "contextual" (in theory), *e.g.*, the terms $\mathcal{O}\phi$ or $\phi*\mathcal{O}\phi$ may show "observation-created" collapse (due to the presence of $\phi$; formalism replacing ontology). Three of the four terms in writing-out Eq. (7) may be "contextual", containing $\phi$ or $\phi*$, and similarly for Eq. (5). The remaining terms however are not "observation-created" in this sense of containing $\phi$ or $\phi*$ : *i.e.,* $|\psi|^2$ in Eq. (5) and $\psi* \mathcal{O}\psi$ in Eq. (7). This point is crucial to our argument that quantum mechanics is not entirely "contextual" (cf. normalization of $\psi + \phi$).

We may redefine (and generalize) Eq. (4), substituting the integrand of Eq. (7), such that upon including $\phi$ we have an observable ensemble average

$$<V> \equiv \int_{\text{all space}} O \, d^3x \quad , \qquad \text{Eq. (8)}$$

integrating over all space ( for position, momentum, or Hamiltonian operators) . Hereinafter $<\mathcal{O}>$ means Eq. (8): This is our new and final interference-collapse equation. Eq. (8) *and* its integrand are being called the collapse consistently. Theory defines stages for us in the theory of measurement, *coupling* ($\psi \otimes \phi \in \mathcal{H}$), *superposition* $\psi + \phi$, *interference*, *via* the cross-terms of O in Eq. (7) , and *collapse* in Eq. (8) for $<V>$. Once collapsed, there is *disturbance* of the system.

In AS, logic and phenomenology develop independently, although in physics an experimentalist may well look at theoretical predictions before setting out on his or her labors. Experiment to date supports quantum mechanics by and large - a full treatment must propose an experiment to find for AS or the Copenhagen interpretation (*i.e.*, non-ontological predictions *vs.* ontology). Firstly, zero-out the pure $\phi$ term (Eqs. 7-8) during calibration with $\psi \equiv 0$. Experiment could test the collapse physics of $\psi* \mathcal{O}\psi$ (Eq. (7)), which in part, when integrated, would give the ensemble average corresponding to $\mathcal{O}$. The total Eq. (8) must yield a subsequent semi-empirical Eq. (4) not involving $\phi$. Eq. (8) must agree with experiment; *cf.* also the term $|\psi|^2$ (Eq.(5)).

Experiment also has something to say about the related problems of the EPR experiment, violation of local realism (Torgerson *et al.* 1995), nonlocality in realistic interpretations of quantum mechanics (Hardy 1992), and entanglement (Ghirardi 2002). It is beyond the scope and purpose of this paper to go into the details of these problems (spacelike acausality). Anyhow, we have a caveat.

So far nobody (including Bohr) has "fully […] worked out a coherent non-ontological interpretation of quantum mechanics", but we may be "forced to accept" one (Shimony 1978, p. 12). This does not include the Copenhagen interpretation; what has been presented in this Section runs counter to the metaphysical (ontological). We have attempted to outline how to proceed, admittedly, with a very simple-minded interpretation example: Whether or not we have succeeded fully, the concept of a manifestly non-ontological interpretation is still a necessary goal for future work. Our present interpretation merely says that a collapse equation exists, and that the foundations of quantum mechanics have compelling aspects of being theoretically physical (*vs.* ontology; "contextual reality" notwithstanding). It has been tried, but perhaps an ensemble approach to quantum mechanics may still be envisaged, even though Ghirardi (2002) says there are no deterministic or random local hidden variables in quantum mechanics (the distributions $\mathcal{P}$ and P are not random).

The AS interpretation is a welcome challenge in this paper, derived by applying the AS rule (no metaphysics) to quantum mechanics. We emphasize that the interpretation example/theory of measurement should not be allowed to detract from AS as worked out and presented in Sect. 2 (in the Table), because it is just an outline of what might be going on in quantum mechanics when we eliminate ontology.

The issues in this Section are fundamental ones when brought to bear on the foundations of quantum mechanics and the Copenhagen interpretation. Finally, to recapitulate, metaphysics plays no role in quantum mechanics using the Logic-Epistemology method.

5. CONCLUSIONS. We identify and define AS as Logic-Epistemology in Physics. Physical Science, the Scientific Method and AS are intimately and basically related; we have compared Theory-Experiment and Hypothesis-Test with Logic-Epistemology in the Table (classically or quantum mechanically).

The theoretical side of AS/logic is constituted by the hypothetical propositions, the reasoning between equations, the equations themselves (formally), and the deduced final result. The logic is incomplete as compared to theory because of the additional mathematics involved (beside logic).

On the experimental side of AS/phenomenology, AS is data-taking and subsequent data-analysis but not earlier stages of experiment such as design, construction and operation (incompleteness). We treat Epistemology as phenomenology and the comparison of phenomenology with logic.

The comparison of logic and phenomenology gives us a law of physics (AS; or it may fail methodologically to confirm a theory, of course). Also, AS takes on the meaning encompassing all the laws in different areas of Physics, when logic and phenomenology agree, and there is no metaphysics.

One of our primary motivations has been to clarify the precise role of philosophy in AS, with its (two) technical branches, the artsciences: logic in physics and epistemology in physics. We see that we must simply retain only necessary and relevant essentials, *i.e.*, Logic-Epistemology, but not metaphysics.

If philosophy is to be used in physics at all, it must be as AS. As long as there are fields like "Philosophy in Physics and Philosophy of Science", to be rigorous, we have to specify what is meant by "Philosophy": It is AS.

Physicists should (and indeed all scientists and philosophers of science may) appreciate the relative simplicity of AS and its ultimate utility in classical *and* quantum physics, and in treating the Copenhagen interpretation. In AS, as compared to Philosophy, we must ignore non-technical-philosophy (metaphysics). There is no possibility of having metaphysical aspects of physical laws or natural phenomena if we follow the AS prescription (Logic-Epistemology) in the amenable cases of classical and quantum physics, each of which have been treated separately (focusing on the Copenhagen interpretation). AS is more natural than Philosophy not containing metaphysical assumptions at all. There is this explicit difference between them; AS qualifies philosophy to be technical-philosophy in physics (the artsciences). AS specifies and exhibits the structure of the artsciences, explicitly as Logic-Epistemology, eliminating metaphysics which might be implicit when one says "Philosophy".

AS falls short of being able to completely model Physical Science. It has been emphasized how the role played by AS is less technical than the Scientific Method (and Physics) clearly showing the limitations of AS. When we say less technical we mean less complete, the role of this incompleteness present in AS being explained in detail. This is interesting in and of itself, having the result that quantum physical phenomenology is similar to the Copenhagen interpretation's considering nothing before measurement (except through theory and statistics).

A non-ontological solution to the "measurement problem" has been sought in a theory of measurement (couple, superpose, interfer, collapse, disturb) subject to experimental verification. We have identified the collapse equation for <V>. The terms P and O describe the collapse of the wavefunction, and $\phi * \mathcal{O} \phi$ (among other terms) may be attributed to "observation-created" collapse due to the presence of $\phi$. Of prime importance (for experiment) are the terms not containing $\phi$. The non-ontological AS interpretation of quantum mechanics is theoretically physical and testable.

In short, we have seen two significant things: 1) what AS does entail (Logic-Epistemology) and what it does not (it is free of metaphysics) and 2) that it is only a partial model of Physical Science (incompleteness),

We can now answer the question that we started out with ... what is AS? AS is a classical or non-ontological quantum mechanical law of physics at the endpoint of any research, by way of the quantitative comparison (agreement) of logic and phenomenology (Logic-Epistemology) - or all such laws of nature and equations in Physics.

Note: There are two References listed (Weinberg 1977a, b) which bear little resemblance to AS as it has evolved and appears in the present paper. They are given just for historical completeness (incidently, the terminology AS was developed independently of the German *Kunstwissenschaft* which is a philosophy of art, as opposed to philosophy of science).

ACKNOWLEDGEMENTS

I would like to thank the University of Rochester for its ongoing support.